\documentclass[english]{llncs}
\ifx\pdfoutput\undefined
\usepackage[dvips]{epsfig}
\else
\usepackage{graphicx}
\fi 
\usepackage{upgreek} 
\usepackage{amsfonts}
\usepackage{amsmath}
\usepackage{amssymb}
\usepackage{bbm}
\usepackage{mathbbol}
\usepackage{mathrsfs}

\renewcommand{\bold}[1]{{\bf #1}}
\newcommand{\sqle}{\sqsubseteq}

\newcommand{\Var}{\mathrm{Var}}

\newcommand{\FV}{\mathrm{FV}}
\newcommand{\subst}[3]{#1[#3/#2]}

\newcommand{\st}{:}

\newcommand{\Mfin}[1]{\mathcal{M}_\mathrm{f}( #1 )}

\newcommand\Omegatuple[1]{\Mfin{#1}^{(\omega)}}

\newcommand{\Env}{\mathrm{Env}}

\renewcommand{\sup}[1]{\bigsqcup #1}

\newcommand{\Id}[1]{\mathrm{Id}_{#1}}
\newcommand{\cat}[1]{\bold{#1}}
\newcommand{\Rel}{\bold{Rel}}

\newcommand{\MRel}{\bold{M\hspace{-1pt}Rel}}
\newcommand{\Termobj}{\mathbbm{1}}
\newcommand{\eval}{\mathrm{eval}}

\newcommand{\proj}[1]{\pi_{#1}}

\newcommand{\Funint}[2]{[{#1}\To{#2}]}
\newcommand{\To}{\Rightarrow}

\newcommand{\lm}[1]{\mathscr{#1}} 
\newcommand{\Lint}[1]{\Lbrack #1\Rbrack} 
\newcommand{\ca}[1]{\mathscr{#1}} 

\newcommand{\Fmor}[2]{\cat{#1}_\mathrm{f}(#2^\Var,#2)}

\newcommand{\ro}[1]{\mathscr{#1}} 
\newcommand{\App}{\mathcal{A}}
\newcommand{\Abs}{\uplambda}
\newcommand{\Cint}[1]{\vert {#1}\vert}

\newcommand{\alg}[1]{\mathscr{#1}} 
\newcommand{\CE}[1]{{\mathrm{CE}(#1)}} 

\newcommand{\LLT}{\lambda\mathscr{T}}
\newcommand{\Th}[1]{\mathrm{Th}(#1)} 
\newcommand{\Thle}[1]{\mathrm{Th_{\sqle}}(#1)} 
\newcommand{\BTth}{\mathcal{B}}

\newcommand{\comp}{\circ}

\newcommand{\cH}{{\cal H}}

\newcommand{\cT}{{\cal T}}

\newcommand{\gth}{\vartheta}

\newcommand{\gl}{\lambda}

\newcommand\ssk{{\bf k}}
\newcommand\sst{{\bf t}}
\newcommand\ssf{{\bf f}}
\newcommand\sss{{\bf s}}
\newcommand\ssK{{\bf K}}
\newcommand\ssS{{\bf S}}

\newcommand{\nat}{\mathbb{N}} 

\title{Models and theories of $\lambda$-calculus}

\author{Giulio Manzonetto\textsuperscript{1,2}\\
        gmanzone@gmail.com
}

\institute{Laboratoire PPS, Universit\'{e} Paris 7,\\
2, place Jussieu (case 7014), 75251 Paris Cedex 05, France
\and
Universit\`a Ca Foscari di Venezia,\\
Dipartimento di Informatica
Via Torino 155,
Venezia, Italy
}

\authorrunning{Manzonetto}

\begin{document}

\maketitle

\begin{abstract}
In this paper we briefly summarize the contents of Manzonetto's PhD thesis \cite{ManzonettoTh}
which concerns denotational semantics and equational/order theories of the pure untyped $\lambda$-calculus.
The main research achievements include: 
$(i)$ a general construction of $\lambda$-models from reflexive objects in (possibly non-well-pointed) categories;
$(ii)$ a Stone-style representation theorem for combinatory algebras; 
$(iii)$ a proof that no effective $\lambda$-model can have $\lambda\beta$ or $\lambda\beta\eta$ as its equational theory 
(this can be seen as a partial answer to an open problem introduced by Honsell-Ronchi Della Rocca in 1984). 

These results, and others, have been published in three conference papers \cite{ManzonettoS06,BerlineMS07,BucciarelliEM07} and 
a journal paper \cite{ManzonettoS08}; a further journal paper has been submitted \cite{BerlineMS08}.

\medskip

{\bf Keywords:} $\lambda$-calculus, lattice of $\lambda$-theories, indecomposable semantics, Stone representation theorem for combinatory algebras,
algebraic incompleteness of $\lambda$-calculus, relational semantics, effective models, L\"owenheim-Skolem theorem for graph models.
\end{abstract}

\section*{Introduction}

The \emph{$\lambda$-calculus} \cite{Church32} was introduced around 1930 by Alonzo Church as the kernel of an investigation in the foundation of mathematics and logic,
where the notion of `function' instead of `set' was taken as primitive.
Subsequently, the $\lambda$-calculus emerged as a consistent fragment of the original system, which became a key tool in the study of computability and, 
with the rise of computers, the formal basis of the functional programming paradigm.
Today, the $\lambda$-calculus plays an important role as a bridge between logic and computer science, which explains the general interest in this formalism 
among computer scientists.

A quarter of century after Barendregt's main book \cite{Bare}, a wealth of interesting problems about models and theories of the untyped $\lambda$-calculus are still open.
In \cite{ManzonettoTh} we use techniques of category theory, universal algebra and recursion theory to shed new light on the known semantics of $\lambda$-calculus,
and on the $\lambda$-theories which can be represented in these semantics. 
Moreover, we introduce and study two new kinds of semantics: the \emph{relational semantics} and the \emph{in\-de\-com\-po\-sable semantics}.

Models of the untyped $\lambda$-calculus may be defined either as reflexive objects in Cartesian closed categories ({\em categorical models} \cite[Sec.~5.5]{Bare})
or as combinatory algebras satisfying some first-order axioms ({\em $\lambda$-models} \cite[Sec.~5.2]{Bare}).

The classic relationships between these two definitions show that these two notions are `almost' equivalent (cf.\  \cite[Ch.~5]{Bare}).
Indeed, given a $\lambda$-model $\ro{C}$ we may build a Cartesian closed category where the underlying set of  $\ro{C}$ is a reflexive object;
conversely, if $U$  is a reflexive object in a {\em well-pointed} Cartesian closed category $\cat C$ then the set $\cat C(\Termobj,U)$ of the `points' of $U$ can be endowed with a structure of $\lambda$-model.
If the category is not well-pointed then the set $\cat C(\Termobj,U)$ cannot be turned into a $\lambda$-model because of the failure of the Meyer-Scott axiom.

In \cite{ManzonettoTh} we provide a new construction which allows to present every categorical model as a $\lambda$-model, even when the underlying category is not well-pointed.
Moreover, we provide sufficient conditions for categorical models living in arbitrary Cartesian closed categories to have as equational theory the greatest sensible consistent $\gl$-theory $\mathcal{H}^*$ (as in the case of Scott's model $\lm{D}_\infty$).
We also build a particularly simple categorical model living in a (highly) non-well-pointed Cartesian closed category of sets and relations (the {\em relational semantics}) 
which satisfies these conditions, and we prove that the associated $\lambda$-model enjoys some algebraic properties which make it suitable for modelling 
non-deterministic extensions of $\lambda$-calculus.

Concerning combinatory algebras, we prove that they satisfy a generalization of Stone's representation theorem stating that every combinatory algebra is 
isomorphic to a weak Boolean product of directly indecomposable combinatory algebras.
The proof of this result involves an application of some non-trivial techniques from universal algebra, thus challenging a general perception that combinatory algebras are 
algebraically `pathological' and hence not particularly amenable to the ideas and methods of universal algebra.

We investigate the semantics of $\lambda$-calculus whose models are directly indecomposable as combinatory algebras, namely the {\em indecomposable semantics}. We show that 
this semantics is large enough to include all the main semantics (i.e., the Scott-continuous, the stable and the strongly stable semantics) and all the term models of semi-sensible $\lambda$-theories. However, we also show that this semantics is largely incomplete, thus giving a {\em new} and {\em uniform} of the incompleteness of the three main semantics.

Finally, we investigate the problem of whether there exists a non-syntactical model of $\lambda$-calculus belonging to the main semantics which has an r.e.\ 
(recursively enumerable) order or equational theory. 
This is a natural generalization of Honsell-Ronchi Della Rocca's longstanding open problem about the existence of a continuous model having as equational theory exactly $\lambda\beta$ or 
$\lambda\beta\eta$ (since $\lambda\beta$ and $\lambda\beta\eta$ are r.e.\ $\lambda$-theories). 
Then, we introduce an appropriate notion of {\em effective model of $\lambda$-calculus}, which covers in particular all the models individually introduced in the literature,
and we prove that no order theory of an effective model can be r.e.; from this it follows that its equational theory cannot be $\lambda\beta$ or $\lambda\beta\eta$.
Furthermore, we show that no effective model living in the stable or strongly stable semantics has an r.e.\ equational theory.
Concerning the continuous semantics, we prove that no order theory of a graph model can be r.e.\ and that many effective graph models do not have an r.e.\ equational theory.\\

{\bf Outline.} In Section~\ref{sec:Preliminaries} we recall some known results concerning the $\lambda$-calculus, and its models and theories.
In Section~\ref{sec:Non-concrete semantics of lambda-calculus} we show that every categorical model of $\lambda$-calculus can be presented as a $\lambda$-model,
and we present a class of categorical models having as theory $\cH^*$.
In Section~\ref{sec:The relational semantics} we build a model $\lm{D}$ living in a relational semantics,
which can be seen as a relational analogue of Scott's $\lm{D}_\infty$.
In Section~\ref{sec:The algebraic incompleteness of gl-calculus} we provide a Stone's representation theorem for combinatory algebras, 
and we study the indecomposable semantics.
Section~\ref{sec:A longstanding open problem, and developments} is devoted to present Honsell-Ronchi Della Rocca's open problem,
and its generalizations.
In Section~\ref{sec:(Concrete) Effective models} we give the notion of effective models of $\lambda$-calculus and 
we study the question of the existence of an effective model having r.e.\ equational/order theories.
In Section~\ref{sec:Graph models: a case of study} we study the class of effective graph models.

\section{Background}\label{sec:Preliminaries}

To keep this paper as self-contained as possible, we summarize some definitions and results used below. 
With regard to the $\gl$-calculus we follow the notation and terminology of \cite{Bare}. 
Our main reference for category theory is \cite{MacLaneS71} and for universal algebra is \cite{BurrisS81}.

\subsection{The $\gl$-calculus}

The two primitive notions of the $\lambda$-calculus are \emph{application}, the operation of applying a function to an argument, and 
\emph{lambda abstraction}, the process of forming a function from the ``expression'' defining it.

The set $\Lambda$ of {\em $\gl$-terms} over a countable set $\Var$ of variables is constructed inductively as follows: 
every variable $x\in\Var$ is a $\lambda$-term; if $M$ and $N$ are $\lambda$-terms, then so are $MN$ and $\lambda x.M$ for each $x\in\Var$.

The lambda abstraction is a binder. An occurrence of a variable $x$ in a $\gl$-term is \emph{bound} if it lies within the scope of a 
lambda abstraction $\gl x$; otherwise it is \emph{free}.
We denote by $\FV(M)$ the set of all free variables of $M$ and we say that $M$ is {\em closed} if $\FV(M)=\emptyset$.
We write $M[N/x]$ for the term resulting by substituting $N$ for all free occurrences of $x$ in $M$ subject to the usual proviso about renaming bound variables
in $M$ to avoid capture of free variables in $N$.

Concerning specific $\lambda$-terms we set (where $\equiv$ denotes syntactical equivalence):
$$
    \ssK\equiv\gl xy.x\qquad \ssS\equiv \gl xyz.xz(yz)
$$

The basic axioms of $\lambda$-calculus are the following:
\begin{itemize}
\item [($\alpha$)] $\lambda x.M = \lambda y.\subst{M}{x}{y}$, for any variable $y$ that does not occur free in $M$;
\item [($\beta$)] $(\lambda x.M)N = \subst{M}{x}{N}$.
\end{itemize}
The rules for deriving equations from instances of ($\alpha$) and ($\beta$) are the usual ones from equational calculus asserting that equality is a
congruence for application and lambda abstraction. 

Extensional $\lambda$-calculus adds another axiom, which equates all the $\lambda$-terms having the same extensional behaviour:
\begin{itemize}
\item[($\eta$)] $\lambda x.Mx = M$, where $x$ does not occur free in $M$.
\end{itemize}

If two $\lambda$-terms are provably equal using the rule ($\alpha$) we say that they are \emph{$\alpha$-equivalent} or \emph{$\alpha$-convertible} 
(and similarly for ($\beta$) and ($\eta$)).
In this work we identify all $\alpha$-equivalent $\lambda$-terms, thus every $\lambda$-term represents in fact a {\em class} of $\alpha$-equivalent terms. 

\subsection{The lattice of $\lambda$-theories}

The \emph{$\lambda$-theories} constitute the main object of study of the untyped $\gl$-calculus when, roughly speaking, we consider the computational equivalence of $\gl$-terms more important than the process of calculus. 
Lambda theories are, by definition, equational extensions of the untyped $\lambda$-calculus which are closed under derivation; more precisely: 
a {\em $\lambda$-theory} is any congruence on the set $\Lambda$ of $\lambda$-terms containing $\alpha$- and $\beta$-conversion \cite[Def.~4.1.1]{Bare}; 
\emph{extensional} $\lambda$-theories are those which contain furthermore the $\eta$-conversion. 
We denote by $\lambda\beta$ ($\lambda\beta\eta$) the least (least extensional) $\lambda$-theory. 

The set of all $\lambda$-theories can be naturally seen as a complete lattice, hereafter denoted by $\LLT$, whose associated ordering is the set-theoretical inclusion.
It is clear that $\lambda\beta$ is the least element of $\LLT$, while the top element is the unique inconsistent $\lambda$-theory $\nabla$ 
({\em inconsistent} because it equates all $\gl$-terms).

Given a $\lambda$-theory $\cT$ we write $M =_\cT N$ if $M,N$ are provable equal in $\cT$.

A $\lambda$-theory $\cT$ is {\em recursively enumerable} ({\em r.e.}, for short) if the set of G\"odel numbers of all pairs of $\cT$-equivalent $\lambda$-terms
is recursively enumerable. For instance, $\lambda\beta$ and $\lambda\beta\eta$ are r.e.\ $\gl$-theories.

Other interesting $\gl$-theories can be defined by classifying $\gl$-terms with respect to their computational behaviour. 
A $\gl$-term $M$ is {\em solvable} if 
$$ M =_{\gl\beta} \gl x_1\dots x_n.x_iM_1\cdots M_k$$ 
for some $n,k\ge 0$ and $M_1,\ldots,M_k\in\Lambda$.
$M$ is {\em unsolvable}, otherwise. 
Intuitively, solvable $\gl$-terms are interesting from the computational point of view since they provide at least a partial fixed output 
(namely, $\gl x_1\dots x_n.x_i-_1\cdots -_k$), 
whilst unsolvable $\gl$-terms correspond to looping terms.
Looking at the $\gl$-theories in terms of solvability/unsolvability, they are classified as \emph{semisensible}, if they do not equate a solvable and an unsolvable $\gl$-term, and as \emph{sensible}, if they are semisensible and they equate \emph{all} unsolvable $\gl$-terms \cite[Sec.~16, 17]{Bare}.
The $\gl$-theory $\cH$, generated by equating all unsolvable $\gl$-terms, is the minimal sensible $\gl$-theory and it is consistent.
$\cH$ admits a unique maximal consistent extension $\cH^*$.

Although researchers have, till recently, mainly focused their interest on a limited number of them, the set of $\lambda$-theories
constitutes a very rich, interesting and complex mathematical structure (see \cite{Bare,Berline00,Berline06}), whose cardinality is $2^{\aleph_0}$.

\begin{center}
\vspace*{1ex}
\ifx\pdfoutput\undefined
\epsfig{file=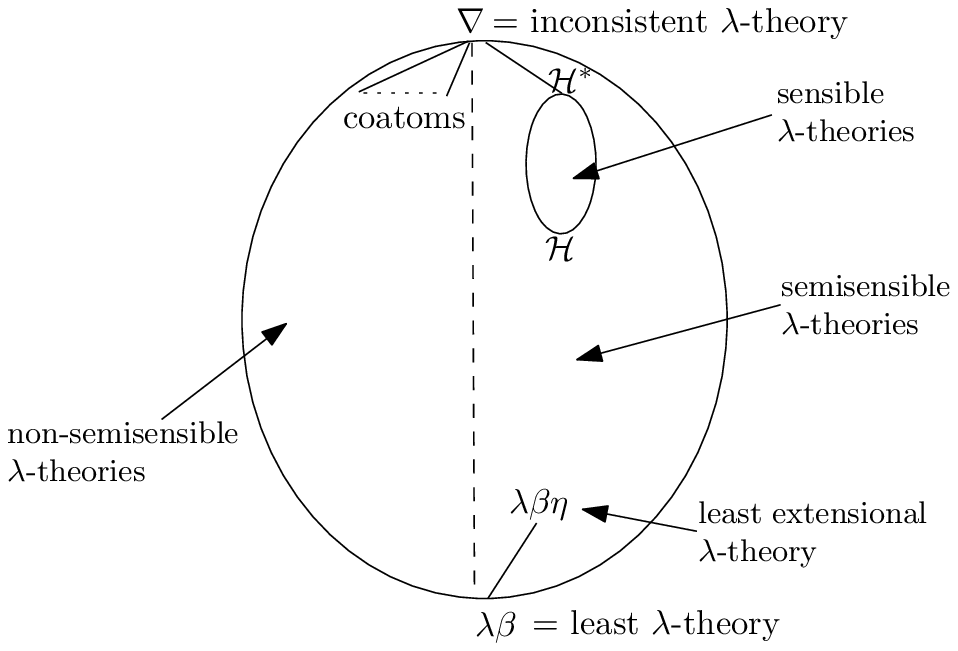}
\else
\includegraphics{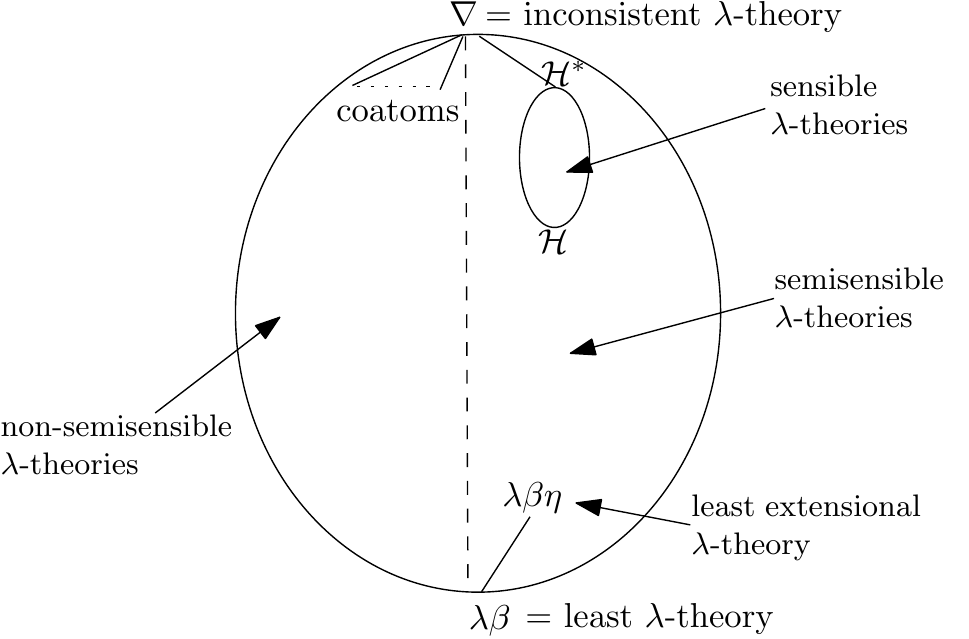}
\fi 
\vspace*{1ex}
\end{center}

Lambda theories arise by syntactical or semantic considerations. 
Indeed, a $\lambda$-theory may correspond to a possible operational (observational) semantics of $\lambda$-calculus, as well as it 
may be induced by a model of $\lambda$-calculus through the kernel congruence relation of the interpretation function.

\subsection{Models of $\lambda$-calculus.} \label{subs:lambda models}

In 1969, Scott found the first model of $\lambda$-calculus ($\ro{D}_\infty$) in the Cartesian closed category (ccc, for short) of complete lattices and 
Scott continuous functions. 
The question of \emph{what is a model of $\lambda$-calculus} has been investigated by several researchers, but only at the end of the seventies
they were able to provide general definitions.
In Manzonetto's PhD thesis \cite{ManzonettoTh} two notions of model of $\lambda$-calculus are mainly used: the former is category-theoretic 
and the latter is algebraic. 

From the categorical point of view, the definition of a model of $\lambda$-calculus is the following. 

\begin{definition} {\em (Categorical model)} A \emph{categorical model} of $\lambda$-calculus is a \emph{reflexive object} of a Cartesian closed category $\cat{C}$, 
i.e., a triple $\ro{U} = (U,\App,\Abs)$ such that $U$ is an object of $\cat{C}$, and  $\App:U\to\Funint UU$ and $\Abs:\Funint UU \to U$
are two morphisms satisfying $\App\comp\Abs=\Id{\Funint{U}{U}}$. 
\end{definition}

The categorical interpretation of a $\lambda$-term $M$ in a model $\ro{U}$ is a morphism $\Cint{M}_I:U^I\to U$ where\footnote{
Here $U^I$ denotes the $I$-indexed categorical product $\prod_{x\in I} U_x$ of several copies of $U$.
} $I$ is some finite subset of $\Var$
containing all the free variables of $M$ (see \cite[Def.~5.5.3(vii)]{Bare}, for a formal definition). 

We write $\ro{U}\models^{\sf categorical} M = N$ if $\Cint{M}_I = \Cint{N}_I$ for some $I\supseteq\FV(MN)$.

From the algebraic point of view, a model of $\lambda$-calculus is a particular kind of combinatory algebra, namely, a {\em $\lambda$-model}.

\begin{definition} 
{\em ($\gl$-model)} A \emph{combinatory algebra} $\ca{C} = (C, \cdot, \ssk, \sss)$ is an applicative structure together with two constants
$\ssk,\sss$ such that 
$$
    \textrm{$\ssk xy=x\quad$ and $\quad\sss xyz=xz(yz)$.}
$$
A combinatory algebra $\ca{C}$ is a {\em $\gl$-model} if, moreover, it satisfies the five axioms of Curry \cite[Thm.~5.2.5]{Bare} and the Meyer-Scott axiom \cite[Def.~5.2.7(ii)]{Bare}.
\end{definition}

The interpretation of $\gl$-terms in a $\gl$-model $\lm{C}$ is defined by using environments:
an {\em environment} with values in $\lm{C}$ is a total function $\rho:\Var\rightarrow\alg{C}$.
We let $\Env_\alg{C}$ be the set of environments with values in $\alg{C}$.

The interpretation of a $\gl$-term $M$ in a $\gl$-model $\lm{C}$ (under an environment $\rho$) is an element of $\lm{C}$ denoted by $\Lint{M}_\rho$ (see \cite[Def.~5.2.1(ii)]{Bare}, for a formal definition).
We write $\ro{C}\models^{\sf algebraic} M = N$ if $\Lint{M}_\rho = \Lint{N}_\rho$ for all $\rho\in\Env_\alg{C}$.

Every {\sf xxx}-model $\lm{M}$ (for {\sf xxx} = categorical/algebraic) induces a $\lambda$-theory 
$$
    \Th{\lm{M}} = \{ M = N \st \lm{M}\models^{\sf xxx} M = N\}
$$
which is called the (equational) theory of $\lm{M}$.

Both {\sf xxx}-notions of model are {\em equationally complete}, in the sense that for every $\gl$-theory $\cT$ there is a {\sf xxx}-model $\lm{M}$ such that $\Th{\lm{M}} = \cT$.
This result is a consequence of the fact that, for each $\lambda$-theory $\cT$, the set of the equivalence classes of $\lambda$-terms modulo 
$\cT$ together with the operation of application on the equivalence classes constitutes a $\lambda$-model $\lm{M}_\cT$ (called the {\em term model of $\cT$}), 
whose theory is exactly $\cT$ \cite[Cor.~5.2.13(ii)]{Bare}.
Moreover, using a construction called {\em Karoubi envelope} \cite[Def.~5.5.11]{Bare}, 
every term model can be seen as a reflexive object in a suitable Cartesian closed category, having the same equational theory.

In the following, we will say that a model of $\lambda$-calculus is \emph{syntactical} if its construction depends on the syntax of $\lambda$-calculus;
in particular, all term models are syntactical.

\subsection{The main semantics}\label{subs:mainsem}

After Scott's $\ro{D}_\infty$, a large number of mathematical models of $\lambda$-calculus, arising from syntax-free constructions, have been introduced 
in various categories of domains and were classified into semantics according to the nature of their representable functions, see e.g.\ \cite{Bare,Berline00,Plotkin93}. 
The {\em continuous semantics} (Scott \cite{Scott72}) is given in the category whose objects are complete partial orders and morphisms are Scott continuous functions. 
The {\em stable semantics} (Berry \cite{Berry78}) and the {\em strongly stable semantics} (Bucciarelli-Ehrhard \cite{BucciarelliE91}) are refinements of the 
continuous semantics, introduced to capture the notion of ``sequential'' Scott continuous function.
By ``the main semantics'' we will understand one of these and, for brevity, we will respectively call the models living inside: 
continuous, stable and strongly stable models.
In each of these semantics all the models come equipped with a partial order, and some of them, called \emph{webbed models}, are built from lower level structures 
called ``webs''.
The simplest class of webbed models is the class of \emph{graph models}, which was isolated in the seventies by Plotkin, Scott and Engeler within the continuous semantics. 

\subsection{The main semantics are incomplete} 

It is well known that the main semantics are {\em equationally incomplete}: 
there are $\lambda$-theories that cannot be represented as theories of models living in the main semantics.
Actually, these semantics do not even match the most natural operational semantics of $\lambda$-calculus.
The problem of the equational completeness was negatively solved by Honsell and Ronchi Della Rocca \cite{HonsellR92} for the 
continuous semantics, by Bastonero and Gouy \cite{BastoneroG99,GouyTh} for the stable semantics and by Bastonero \cite{BastoneroTh} for the strongly stable semantics.
In \cite{Salibra01,Salibra03} Salibra proved that the ``monotonous semantics'' (i.e., the class of all $\lambda$-models involving monotonicity with 
respect to some partial order and having a bottom element) is incomplete, thus giving a first uniform proof of incompleteness encompassing the three 
main semantics.


\section{Non-concrete semantics of $\lambda$-calculus}\label{sec:Non-concrete semantics of lambda-calculus}

In \cite{ManzonettoTh} we are mainly interested in models of $\lambda$-calculus living in the main semantics. 
The ccc's underlying the main semantics, are all {\em well-pointed} which means that for all objects $A,B$ and 
morphisms $f,g:A\to B$, whenever $f\neq g$, there exists a morphism $h:\Termobj\to A$, such that $f\comp h \neq g\comp h$. 
In other words, the category has enough `points' to separate all different morphisms.

However, in denotational semantics, non-well-pointed ccc's arise naturally when morphisms are not functions, as e.g., 
sequential algorithms \cite{BerryC85} or strategies in various categories of games \cite{AbramskyJM00,HylandO00}. 
In \cite[Chapter~3]{ManzonettoTh} we build and study a categorical model $\ro{D}$ living in a (highly) non-well-pointed ccc of 
sets and relations. 

If we choose as definition of model of $\lambda$-calculus the notion of $\lambda$-model, we could be reluctant to consider $\ro{D}$ as a real model, 
since the only known construction for turning a categorical model into a $\lambda$-model asks for well-pointed ccc's (see, e.g., \cite[Ch.~5]{Bare}).
In the next subsection we will see that this fact does not constitute a problem: 
indeed, in \cite[Chapter~2]{ManzonettoTh} we describe an alternative construction which works in greater generality and allows us to present 
{\em any} categorical model as a $\lambda$-model.

\subsection{From Cartesian closed categories to $\lambda$-models}

It is well known that the notions of $\lambda$-model and of categorical model are equivalent in the following sense (see \cite[Ch.~5]{Bare} and \cite[Sec.~9.5]{AspertiL91}).
Given a $\lambda$-model $\ca{C}$ we may build a ccc in which the underlying set of $\alg{C}$ is a reflexive object; 
conversely, if $\ro{U}$ is a reflexive object in a well-pointed ccc $\cat{C}$, then the set\footnote{
$\cat{C}(\Termobj,U)$ denotes the homset of morphisms of the form $f:\Termobj\to U$.
} $\cat{C}(\Termobj,U)$ of the `points' of $U$ 
can be endowed with a structure of $\lambda$-model.

If the ccc is not well-pointed then, in general, $\cat{C}(\Termobj,U)$ cannot be turned into a $\lambda$-model, because of the failure of the Meyer-Scott axiom.
In the next theorem we show that this apparent mismatch can be avoided by changing $\cat{C}(\Termobj,U)$
for the set $\Fmor{C}{U}$ of `finitary' morphisms in $\cat{C}(U^\Var,U)$. 
Intuitively, a morphism $f:U^\Var\to U$ is {\em finitary} if it only depends on a finite number of variables, i.e., if it can be decomposed as $f = f_I\comp \proj{I}^\Var$
where $\proj{I}^\Var:U^\Var\to U^I$ is the canonical projection and $I\subset\Var$ is finite. 


\begin{theorem} \cite[Thm.~2.2.12]{ManzonettoTh} Let $\ro{U} = (U,\App,\Abs)$ be a reflexive object living in a (possibly non-well-pointed) ccc $\cat{C}$. 
Then $\alg{C}_{\ro{U}} = (\Fmor{C}{U}, \bullet)$ where $\bullet = \eval\comp\App\times\Id{}$ can be endowed with a structure of $\gl$-model.
\end{theorem}

Using this construction, we can easily switch from the categorical to the algebraic interpretation of $\lambda$-terms and {\em vice versa}.
Actually, the interpretation of a $\gl$-term $M$ (under an environment $\rho:\Var\to\Fmor{C}{U}$) is defined in terms of the categorical interpretation 
of $M$ in $\ro{U}$ as follows:
$$
\Lint{M}_\rho = \Cint{M}_I \comp \langle\rho(x_i)\rangle_{x_i\in I} \textrm{ for some finite $I\supseteq\FV(M)$.}
$$
Conversely, by using a particular environment $\hat{\rho}$, defined by $\hat{\rho}(x) = \proj{x}^\Var$ for all $x\in\Var$,
it is possible to ``recover'' the categorical interpretation 
$\Cint{M}_I$ from the interpretation $\Lint{M}_\rho$. Indeed, we have that:
$$
    \Lint{M}_{\hat\rho}=\Cint{M}_I\comp \pi_I^\Var,
$$
i.e., $\Lint{M}_{\hat\rho}$ is the morphism $\Cint{M}_I$ ``viewed'' as an element of $\cat{C}(U^{\Var},U)$.

It is of course important to notice that our construction, as well as the classical one, preserves the equalities between the denotations of the $\lambda$-terms.
This means that two $\lambda$-terms have the same interpretation in a categorical model if, and only if, they have the same interpretation in the associated 
$\lambda$-model (thus $\Th{\ro{U}} = \Th{\alg{C}_{\ro{U}}}$).


\subsection{A class of models of $\cH^*$}

The $\lambda$-theory $\cH^*$, which is the unique maximal consistent sensible $\lambda$-theory, was first introduced by 
Hyland \cite{Hyland75} and Wadsworth \cite{Wadsworth76}, who proved (independently) that $\cH^*$ is the theory of 
Scott's $\ro{D}_\infty$ \cite[Thm.~19.2.9]{Bare}.
Today, it is well known that there exist several models of $\lambda$-calculus having as equational theory $\cH^*$
\cite{Bare,GouyTh,DiGianantonioFH91}.
The most general result in this context is in Gouy's PhD thesis \cite{GouyTh}. 
Gouy introduces a notion of ``regular ccc'' and characterizes a class of models, all living in regular ccc's, 
which can be suitably stratified yielding $\cH^*$ as equational theory. 
However, regular ccc's are all well-pointed by definition.

In \cite[Chapter~2]{ManzonettoTh} we generalize this result in order to cover also models living in non-well-pointed ccc's. 
The idea is to find a class of models, as large as possible, satisfying a (strong) Approximation Theorem.
More precisely, we want to be able to characterize the interpretation of a $\lambda$-term $M$ (that here is an element of $\cat C(U^\Var,U)$) 
as the least upper bound of its approximants.

Thus, it is natural to ask that the set $\cat C(U^\Var,U)$ constitutes a complete partial order (cpo, for short) 
and that every morphism $a:U^\Var\to U$ has a directed set of approximants $a_k$'s (for $k\in\nat$).
In the following theorem, we indicate by $(a)_k$ the $k$-th approximant of $a:U^\Var\to U$.


\begin{theorem}\label{thm: a model of H*}\cite[Thm.~2.3.35]{ManzonettoTh}
Let $\ro{U} = (U,\App,\Abs)$ be an extensional\footnote{
We recall that a categorical model $\ro{U}$ is {\em extensional} if $U\cong\Funint UU$.
This condition holds exactly when $\lambda\beta\eta\subseteq\Th{\ro{U}}$.} categorical model living in a ccc $\cat C$.
If $\cat C$ is {\em cpo-enriched}, i.e., if every homset has a structure of cpo
$$
    (\cat C(A,B),\sqle,\bot)
$$
and if, for all $a : U^\Var\to U$, the following conditions are satisfied (for all $k\in\nat$)
$$
\begin{array}{lcl}
\bot a = \bot,         & \qquad \qquad \qquad & a_k\sqle a,\\
\Abs\comp \bot = \bot, &        & \sup a_k = a,\\
a_0b = (a\bot)_0,      &        & a_{k+1}b = (ab_k)_k.\\
\end{array}
$$
then $\Th{\ro{U}} = \cH^*$.
\end{theorem}

This generalization is necessary to cover the categorical model $\ro{D}$ built in \cite[Chapter~2]{ManzonettoTh},
whose construction is recalled in the next section.


\section{The relational semantics}\label{sec:The relational semantics}

In \cite[Chapter~3]{ManzonettoTh}  we present a non-well-pointed semantics much simpler than game semantics. 
The category we have in mind is $\MRel$, the Kleisli-category of the comonad of ``finite multisets'' over the category $\Rel$ of sets and relations.
This category can be described explicitely as the category whose objects are sets and whose morphisms $f:S\to T$ are relations $\Mfin{S}\times T$, 
where $\Mfin S$ is the set of finite multisets over $S$.
$\MRel$ is a ccc which has been studied in particular as a semantic framework for linear logic \cite{Girard88,BucciarelliE01};
here, we study it as a {\em relational semantics} of $\lambda$-calculus.

\subsection{A relational analogue of $\ro{D}_\infty$}

In this section we build a reflexive object $\ro{D} = (D,\App,\Abs)$ of $\MRel$, which satisfies $D\cong \Funint DD$ by construction 
(and this implies that $\ro{D}$ is extensional). 

Given a set $A$, we write $\Omegatuple{A}$ for the set of infinite sequences $(m_1,m_2,\ldots)$ of finite multisets over $A$,
such that $m_i$ is non-empty only for a finite number of indices $i$.

By mimicking the classical construction used for graph models, we define an increasing family of sets $(D_n)_{n\in\nat}$ as follows: 
\begin{itemize}
\item $D_0=\emptyset$,
\item $D_{n+1}=\Omegatuple{D_n}$.
\end{itemize}
Finally, we set $D=\cup_{n\in\nat}D_n$.

The key remark to define an isomorphism in $\MRel$ between $D$ and $\Mfin{D}\times D$ (which represents $\Funint{D}{D}$ in $\MRel$) 
is that every element $(\sigma_0, \sigma_1, \sigma_2,\ldots)\in D$ is canonically associated with the pair $(\sigma_0, (\sigma_1, \sigma_2,\ldots))\in \Mfin{D}\times D$ and {\em vice versa}. 
Thus, our categorical model $\ro{D}$ is just the set $D$, endowed with the morphisms $\App:D\to\Funint DD$ and $\Abs:\Funint DD\to D$ performing 
the above bijection.

It is not difficult to check that the category $\MRel$ is cpo-enriched and that 
we can define the $k$-th approximant $a_k$ of a morphism $a:D^\Var\to D$ by using the natural stratification of $D$.
Hence, the following theorem is just a corollary of Theorem~\ref{thm: a model of H*}

\begin{theorem} \cite[Cor.~3.3.4]{ManzonettoTh} $\Th{\ro{D}} = \cH^*$.
\end{theorem}

Finally, in \cite[Chapter~3]{ManzonettoTh} we prove that the $\lambda$-model $\ca{C}_\ro{D}$ associated with $\ro{D}$ by our construction
has a rich algebraic structure.
In particular, we define two operations of sum and product which are left distributive with respect to application and 
give to $\ca{C}_\ro{D}$ a structure of commutative semiring. 
This opens the way to the interpretation of conjunctive-disjunctive $\lambda$-calculi (see, e.g., \cite{DezaniLP98}) 
in this relational framework.

\section{The algebraic incompleteness of $\gl$-calculus}\label{sec:The algebraic incompleteness of gl-calculus}

\subsection{The $\lambda$-calculus does not look algebraic}

The $\lambda$-calculus is not a genuine equational theory since the variable-binding properties of lambda abstraction prevent 
variables in $\lambda$-calculus from playing the role of real algebraic variables.
Consequently, the general methods that have been de\-ve\-lo\-ped in universal algebra are not directly applicable.

There have been several attempts to reformulate $\lambda$-calculus as a purely algebraic theory.

The earliest algebraic models are the $\lambda$-models recalled in Subsection~\ref{subs:lambda models},
which provide a first-order (but not equational) characterization of the models of $\lambda$-calculus.
Recently, Pigozzi and Salibra proposed the \emph{Lambda Abstraction Algebras} as an alternative first-order 
description of the models of $\lambda$-calculus \cite{PigozziS98,Salibra00}.
Lambda Abstraction Algebras form an equational class and allow to keep the lambda-notation and, hence, all the functional intuitions. 

Combinatory algebras are considered ``bad algebras'' since they are never commutative, associative, finite or recursive \cite[Prop.~5.1.15]{Bare}.
Moreover, Salibra and Lusin \cite{LusinS04} showed that only trivial lattice identities are satisfied by all congruence lattices of combinatory algebras. 
Thus, it is not possible to apply in this context the nice results developed in universal algebra in the last thirty years 
(see, e.g., \cite{BurrisS81,McKenzieMT87}) which connect the lattice identities satisfied by all the congruence lattices of algebras belonging 
to a variety, with Mal'cev conditions.

These negative results led to a common belief stating that $\lambda$-calculus and combinatory algebras are algebraically pathological.
As we will see in the next subsection this belief is false: indeed, combinatory algebras do satisfy a Stone-like representation theorem 
which has interesting consequences in the study of $\lambda$-calculus.

\subsection{A Stone's Representation Theorem for combinatory algebras}

One of the milestones of modern algebra is the Stone representation theorem for Boolean algebras.
This result was first generalized by Pierce to commutative rings with unit and next by Comer to the class of algebras with Boolean factor 
congruences (see \cite{Comer71,Johnstone82,Pierce67}). 
By applying a theorem due to Vaggione \cite{Vaggione96b}, we show that Comer's generalization holds for combinatory algebras: 
any combinatory algebra is isomorphic to a weak Boolean product of directly indecomposable combinatory algebras 
(i.e., algebras which cannot be decomposed as the Cartesian product of two other non-trivial algebras). 

The proof of the representation theorem for combinatory algebras is based on the fact that every combinatory algebra has 
{\em central elements}, i.e., elements which induce a direct decomposition of the algebra as the Cartesian product of two 
other combinatory algebras, just like idempotent elements in rings or complemented elements in bounded distributive lattices.

Central elements, which have been introduced by Vaggione \cite{Vaggione96a} in the context of Universal Algebra, admit in the 
context of combinatory algebras the following (elegant) equational characterization (where $\sst \equiv \ssk$ and 
$\ssf \equiv \ssk(\sss\ssk\ssk)$ are the combinatory terms representing the Boolean values).

\begin{definition} An element $e$ of a combinatory algebra $\alg{C}$ is {\em central} if:
\begin{enumerate}
\item[$(i)$] $exx = x$.
\item[$(ii)$] $e(exy)z = exz = ex(eyz)$.
\item[$(iii)$] $e(xy)(zt) = exz(eyt)$.
\item[$(iv)$] $e = e\sst\ssf$.
\end{enumerate}
\end{definition}

Every central element $e$ induces a decomposition $\alg{C}\cong \alg{C}/\gth(e,\sst)\times\alg{C}/\gth(e,\ssf)$, where 
$\gth(a,b)$ denotes the congruence on $\alg{C}$ generated by equating $a$ and $b$.
So, the existence of central elements is just another way of saying ``this algebra is decomposable as a product of simpler algebras''.

In the next theorem we show that the central elements of a combinatory algebra constitute a Boolean algebra, whose Boolean operations
can moreover be internally defined by suitable combinatory terms. 

\begin{theorem} \cite[Thm.~4.3.6]{ManzonettoTh} For every combinatory algebra $\alg{C}$, the set $\CE{\alg{C}}$ of its central elements endowed with the operators:
$$
   e\lor d = ed\ssf; \quad e\land d = e\sst d;\quad e^- = e\ssf\sst,
$$
is a Boolean algebra. 
\end{theorem}

The above result suggests a connection between propositional classic logic and combinatory logic,
and allows us to prove the promised representation theorem.

\begin{theorem} \cite[Thm.~4.3.7]{ManzonettoTh} {\em (Stone's representation theorem for combinatory algebras)}
Every combinatory algebra $\ca{C}$ can be represented as a {\bf weak} Boolean product of indecomposable combinatory algebras $\ca{C}_x$ (for $x\in \alg{X}$):
$$
    \ca{C} \cong \prod_{x\in \alg{X}} \ca{C}_x
$$
where $\alg{X}$ is the Boolean space of maximal ideals of the Boolean algebra $\CE{\alg{C}}$.
\end{theorem}

Note that this result cannot be improved, in the sense that - in general - it is not possible to obtain a (non-weak) Boolean product 
representation of a combinatory algebra (this follows from two results due to Vaggione \cite{Vaggione96a} and Plotkin-Simpson \cite[Lemma~3.14]{SelingerTh}).

\subsection{The indecomposable semantics} 

The representation theorem of combinatory algebras can be roughly summarized as follows: 
the directly indecomposable combinatory algebras are the ``building blocks'' in the variety of combinatory algebras.
Then it is natural to investigate the class of models of $\lambda$-calculus, which are directly indecomposable
as combinatory algebras. We call this class the {\em indecomposable semantics}.

It turns out that the property of ``being an indecomposable model'' is very natural, indeed
the indecomposable semantics is huge enough to encompass the three main semantics. 

\begin{theorem}\cite[Cor.~4.4.9]{ManzonettoTh} The indecomposable semantics includes the continuous, the stable and the strongly stable semantics, as well as the term models 
of all semi-sensible $\lambda$-theories.
\end{theorem}

Moreover, the indecomposable semantics include also all term models of semi-sensible $\lambda$-theories.

\begin{theorem}\cite[Thm.~4.4.7]{ManzonettoTh} If $\cT$ is semi-sensible, then the term model $\lm{M}_\cT$ is indecomposable.
\end{theorem}

However, we also prove that the indecomposable semantics is incomplete, and that this incompleteness is, also, as large as possible: 
\begin{theorem} \cite[Thm.~4.4.16, Thm.~4.4.14]{ManzonettoTh} \
\begin{itemize}
\item[$(i)$] there exists a continuum of pairwise incompatible $\lambda$-theories which are omitted by the indecomposable semantics;
\item [$(ii)$] for every r.e.\ $\lambda$-theory $\cT$ there is a continuum of $\lambda$-theories 
               including $\cT$, and forming an interval, which are omitted by the indecomposable semantics. 
\end{itemize}
\end{theorem}

This gives a \emph{new} and uniform proof of the large incompleteness of each of the main semantics.

In one of the last results of \cite[Chapter~4]{ManzonettoTh} we also show that the set of $\lambda$-theories 
representable in each of the classic semantics of $\lambda$-calculus is not closed under 
finite intersection; hence in particular it is not a sublattice of $\LLT$.

%
%

\section{A longstanding open problem, and developments}\label{sec:A longstanding open problem, and developments}

\subsection{The initial problem}

The question of the existence of a continuous model or, more generally, of a non-syntactical model of $\lambda\beta$ (or $\lambda\beta\eta$) 
has been circulating since at least the beginning of the eighties\footnote{See Problem 22 in the list of TLCA open problems \cite{HonsellTLCA}.}, 
but it was only first raised in print in a paper by Honsell-Ronchi Della Rocca \cite{HonsellR92}.
This problem is still open, but generated a wealth of interesting research and results (surveyed in \cite{Berline00} and \cite{Berline06}). 

In 1995 Di Gianantonio, Honsell and Plotkin succeeded to build an extensional model having theory $\lambda\beta\eta$, living in some 
weakly continuous semantics \cite{DiGianantonioHP95}. 
However, the construction of this model as an inverse limit starts from the term model of $\lambda\beta\eta$, and hence involves the 
syntax of $\lambda$-calculus. 
Furthermore, the problem of whether there exists a model of $\lambda\beta$ or $\lambda\beta\eta$ {\em living in one of the main semantics} 
remains completely open. 
Nevertheless, the authors also proved in \cite{DiGianantonioHP95} that the set of extensional theories representable by models living in
Scott's semantics had a least element. 
At the same time Selinger proved that if an ordered model has theory $\lambda\beta$ or $\lambda\beta\eta$ then the order must be discrete on the interpretations of closed $\lambda$-terms \cite{Selinger03}.

In view of the second result of \cite{DiGianantonioHP95}, it becomes natural to ask whether, given a (uniformly presented) class of models of $\lambda$-calculus, 
there is a minimum $\lambda$-theory represented in it; a question which was raised in \cite{Berline00}. 
In \cite{BucciarelliS04,BucciarelliS0X} Bucciarelli and Salibra showed that the answer is positive for the class of \emph{graph models}, and that 
the least \emph{graph theory} (theory of a graph model) was different from $\lambda\beta$ and 
$\lambda\beta\eta$.
At the moment the problem remains open for the other classes of models.

Ten years ago, it was proved that in each of the known (uniformly presented) classes of models, living in any of the above mentioned semantics, 
and to begin with the class of graph models, it is possible to build $2^{\aleph_{0}}$ (webbed) models inducing pairwise distinct $\lambda$-theories 
\cite{Kerth98b,Kerth01}. 
More recently, it has been proved in \cite{Salibra03} that there are $2^{\aleph_{0}}$ $\gl$-theories which are omitted by all these classes, among which 
$\aleph_{0}$ are finitely axiomatizable over $\lambda\beta$.

From these results, and since there are only $\aleph_{0}$ r.e.\ $\gl$-theories, it follows that each class represents $2^{\aleph_{0}}$ non r.e.\ 
$\gl$-theories and omits $\aleph_{0}$ r.e.\ $\gl$-theories. 
Note also that there are only very few theories of non-syntactical models which are known to admit an alternative description 
(e.g. via syntactical considerations), and that all happen to coincide either with the theory $\BTth$ of B\"{o}hm trees \cite[Sec.~16.4]{Bare} 
or some variations of it, and hence are not r.e. 

This leads us to raise the following problem, which is a second natural generalization of the initial problem.

\subsection{Can a non-syntactical model have an r.e.\ theory?} 

In \cite[Chapter~6]{ManzonettoTh} we investigate the problem of whether the equational theory of a non-syntactical model of $\lambda$-calculus living in one of the main semantics can be r.e.\ 
(note that this is actually a generalization of Honsell-Ronchi Della Rocca's open problem since $\lambda\beta$ and $\lambda\beta\eta$ are r.e.).
As far as we know, this problem was first raised in \cite{Berline06}, where it is conjectured that no graph model can have an r.e.\ equational theory,
but we expect that this could indeed be true for all models living in the continuous semantics, or in its refinements 
(but of course not in its weakenings, because of \cite{DiGianantonioHP95}), and in \cite{ManzonettoTh} this conjecture is officially extended:

\begin{conjecture}\label{conj:ScottSemantics} 
No model living in Scott's continuous semantics or in one of its refinements has an r.e.\ equational theory.
\end{conjecture}

In order to investigate this conjecture, we adopt the following metodology:\\
1) {\em We look also at order theories.} Since all the models we are interested in are partially ordered, and since, in this case, 
the equational theory $\Th{\lm{M}}$ is easily expressible from its order theory\footnote{
In particular, if $\Thle{\lm{M}}$ is r.e.\ then also $\Th{\lm{M}}$ is r.e.
}
$$
    \Thle{\lm{M}} = \{ M \sqle N \st \Lint{M}^\lm{M}_\rho \sqle_{\lm{M}} \Lint{N}^\lm{M}_\rho \textrm{ for all }\rho\in\Env_{\lm{M}}\}
$$ 
we also address the analogue problem for order theories.\\
2) {\em We look at models with built-in effectivity properties}. 
There are several reasons to do so. 
First, it may seem reasonable to think that, if effective models do not even succeed to have an r.e.\ theory, then it is
unlikely that the other ones may succeed; second, because all models which have been individually studied or given as examples in the literature are effective, in our sense.
Starting from the known notion of an effective domain, we introduce an appropriate notion of an \emph{effective model of $\lambda$-calculus}
and we study the main properties of these models. 
Note that, in the absolute, effective models happen to be rare, since each ``uniform'' class represents
$2^{\aleph_{0}}$ $\gl$-theories, but contains only $\aleph_{0}$ non-isomorphic effective models! 
However, and this is a third \emph{a posteriori} reason to work with them, it happens that they can be used to prove 
properties of non effective models (Theorem \ref{Theorem graph} below is the first example we know of such a result).\\
3) A previous result obtained for \emph{typed $\lambda$-calculus} also justifies the above metho\-do\-logy. 
Indeed, it was proved in \cite{BerardiB02} that there exists a (webbed) model of Girard's system $F$, 
living in the continuous semantics, whose theory is the typed version of $\lambda\beta\eta$, and whose
construction does not involve the syntax of $\lambda$-calculus. 
Furthermore, this model can easily be checked to be ``effective'' in the same spirit as in
the present work (see \cite[Appendix C]{BerardiB02} for a sketchy presentation of the model). 
Note that this model has no analogue in the stable semantics.\\
4) \emph{We look at the class of graph models}. 
In order to attack difficult open problems as the one stated above, it is often convenient to focus the attention, first, on the class of graph models.
Indeed, for all webbed models it is possible to infer properties of the models by analyzing the structure of their web, 
and graph models have the simplest kind of web.
Moreover, the class of graph models is very rich, since it represents $2^{\aleph_{0}}$ pairwise distinct (non extensional) $\lambda$-theories. \\
%
5) \emph{We prove a L\"owenheim-Skolem theorem}. 
Effective webbed models are, in particular, generated by countable webs. 
A key step for attacking the general conjecture is hence to prove that the order/equational theory of any webbed model 
can be represented by a model of the same kind but having a countable web. 
In \cite[Chapter~5]{ManzonettoTh} we have proved this theorem for graph models; this result is presented below as Theorem~\ref{LSKgen}.

\section{(Concrete) Effective models versus r.e.\ $\lambda$-theories}\label{sec:(Concrete) Effective models}

As recalled in Subsection~\ref{subs:mainsem}, the models of $\lambda$-calculus living in the main semantics have as underlying sets particular 
cpo's called ``domains''.
The standard notion of computability cannot be directly applied to domains: a trivial reason is that many domains of interest are uncountable.
These domains are conceived as the {\em completion} of a countable set of concrete elements (the compact elements) and computations
on an element in the completion are determined by the way the computations act on its approximations (the compact elements below it).
The theory of computability on domains reflects this idea in the sense that a domain is \emph{effective} when an effective numeration of its 
compact elements is provided \cite[Ch~10, Def.~3.1]{Viggo94}.

Starting from the well-established notion of an effective domain, we introduce an appropriate notion of an {\em effective model of $\lambda$-calculus}
which covers in particular all the models individually introduced in the literature.

\begin{definition} Let $\lm{M}$ be a model living in the continuous semantics.
\begin{itemize}
\item[(i)] $\lm{M}$ is {\em weakly effective} if it is a reflexive object in the category having effective domains as objects and 
computable continuous functions as morphisms. 
\item[(ii)] $\lm{M}$ is {\em effective} if it satisfies a further technical condition implying that the interpretation of every
closed normal $\lambda$-term in $\lm{M}$ is a `decidable' element (see \cite[Def.~6.2.45]{ManzonettoTh} for more details).
\end{itemize}
\end{definition}

The above definitions extend to stable (resp.\ strongly stable) models by taking, as underlying category, the category having effective 
dI-domains (resp.\ dI-domains with coherences) as objects and computable stable functions (resp.\ computable strongly stable functions) as morphisms.

The central technical devices that we use to obtain the results below are:
$1)$ Visser's result \cite{Visser80} stating that the complements of $\beta$-closed r.e.\ sets of $\lambda$-terms enjoy the finite intersection property; 
$2)$ Selinger's theorem stating that every model of $\lambda\beta$ or $\lambda\beta\eta$ must be discretly ordered on the interpretations 
of closed $\lambda$-terms \cite{Selinger03}. 

Our main result in this context is the following.

\begin{theorem} \cite[Cor.~6.2.47, Cor.~6.2.49, Prop.~6.2.40, Thm.~6.2.44]{ManzonettoTh}\\
Let $\lm{M}$ be an \emph{effective} model of $\lambda$-calculus. Then:
\begin{itemize}
\item[(i)] $\Thle{\lm{M}}$ is not r.e.,
\item[(ii)] $\Th{\lm{M}}\neq\lambda\beta,\lambda\beta\eta$,
\item[(iii)] If $\bot_{\lm{M}}$ is $\lambda$-definable then $\Th{\lm{M}}$ is not r.e., more generally:
\item[(iv)] If there is a $\lambda$-term $M$ which is interpreted in $\lm{M}$ as a decidable element having only finitely many 
$\lambda$-definable elements below it, then $\Th{\lm{M}}$ is not r.e.
\end{itemize}
\end{theorem}

Concerning the existence of a non-syntactical effective model with an r.e.\ equational theory, we are able to give a definite answer for all effective stable and strongly stable models:

\begin{theorem} \cite[Thm.~6.3.2]{ManzonettoTh}
No effective model living in the stable or in the strongly stable semantics has an r.e.\ equational theory.
\end{theorem}

This theorem solves Conjecture~\ref{conj:ScottSemantics} for effective models living in these two semantics. 
Concerning Scott's continuous semantics, the problem looks much more difficult and we concentrate on the class of graph models.

\section{Graph models: a case of study}\label{sec:Graph models: a case of study}

In \cite[Chapter~5]{ManzonettoTh} we recall a free completion process for building (the web of) a graph model starting from a ``partial web'',
and we develop some mathematical tools for studying the framework of partial webs. 
These tools are then fruitfully applied to prove the following results.

\begin{theorem}\label{Theorem min} \cite[Thm.~6.4.22]{ManzonettoTh}
There exists an \emph{effective} graph model whose equational/order theory is the minimum graph theory.
\end{theorem}

\begin{theorem}\label{Theorem graph} \cite[Thm.~6.4.24]{ManzonettoTh}
If $\lm{G}$ is a graph model then $\Thle{\lm{G}}$ is not r.e.
\end{theorem}

We emphasize that Theorem \ref{Theorem graph}, which happens to be a consequence of Theorem~\ref{Theorem min}, 
plus the work on effective models, concerns all the graph models and not only the effective ones.
Concerning the equational theories of graph models we only give below, as Theorem~\ref{Theorem graph2}, the more flashy example of the results 
we have proved in \cite[Subsection~6.4.3]{ManzonettoTh}. 
The stronger versions are however natural, and needed for covering all the traditional models.

\begin{theorem}\label{Theorem graph2} \cite[Thm.~6.4.11]{ManzonettoTh}
If $\lm{G}$ is a graph model which is ``freely generated from a finite partial web'', then $\Th{\lm{G}}$ is not r.e.
\end{theorem}

It remains open whether the minimum equational graph theory is r.e.
Hence, the following instances of Conjecture~\ref{conj:ScottSemantics}  are still open; we state them from the weaker to the stronger one.


\begin{conjecture}
The minimum equational graph theory is not r.e.
\end{conjecture}

\begin{conjecture}
All the effective graph models have non r.e.\ equational theories.
\end{conjecture}

\begin{conjecture}
All the effective models living in the continuous semantics have non r.e.\ equational theories.
\end{conjecture}


The following further theorem states that graph models with countable webs are enough for representing all graph theories.
This can be viewed as a kind of L\"{o}wenheim-Skolem Theorem for graph models.

\begin{theorem}\label{LSKgen} \cite[Thm.~5.3.2]{ManzonettoTh} For any graph model $\lm{G}$ there is a graph model $\lm{G'}$ which has a countable web and the same order theory 
(and hence the same equational theory).
\end{theorem}

This result answers positively Problem 12 in \cite{Berline06}.


\section*{Conclusions}

Although Alonzo Church introduced the untyped $\lambda$-calculus in the thirties, the study of its models and theories is, today, 
a research field which is still full of life.
The wealth of results which have been discovered in the last years allow us to understand much better the known semantics of $\lambda$-calculus 
and the structure of the lattice of $\lambda$-theories, but they also generated a lot of new interesting open questions.

In \cite{ManzonettoTh} we have mainly focused our attention on the models of $\lambda$-calculus living in the main semantics, 
but we have also studied two new kinds of semantics: the {\em relational semantics} and the {\em indecomposable semantics}.

Since the relational semantics is non-well-pointed, and ``well-pointedness'' is advocated in the literature as necessary
to obtain a $\lambda$-model, we have found it natural to reinvestigate, first, the relationship between the categorical and algebraic definitions of 
model of $\lambda$-calculus. 
In \cite[Chapter~2]{ManzonettoTh} we have given a new construction which allows us to present {\em any} categorical model 
as a $\lambda$-model, and hence proved that there is a {\em unique} definition of model of $\lambda$-calculus. 
Moreover, we have provided sufficient conditions for categorical models living in arbitrary cpo-enriched ccc's to have $\cH^*$
as equational theory.

In \cite[Chapter~3]{ManzonettoTh} we have built a categorical model $\ro{D}$ living in the relational semantics,
and we have proved that its equational theory is $\cH^*$ since it fulfills the conditions described in \cite[Chapter~2]{ManzonettoTh}.
Then, we have applied to $\ro{D}$ our construction and shown that the associated $\lambda$-model satisfies suitable algebraic properties for modelling a $\lambda$-calculus
extended with both non-deterministic choice and parallel composition.

Concerning the indecomposable semantics, we have proved in \cite[Chapter~4]{ManzonettoTh} that it encompasses the main semantics, as well as the term models of all 
semi-sensible $\lambda$-theories and that, however, it is still largely incomplete.
This gives a new and shorter common proof of the (large) incompleteness of the continuous, stable, and strongly stable semantics.

In \cite[Chapter~5]{ManzonettoTh} we have developed some mathematical tools for studying the framework of partial webs of graph models. 
These tools have been fruitfully used to prove, for example, that there exists a minimum order/equational graph theory and that graph models enjoy a kind of 
L\"owenheim-Skolem theorem.

Finally, in \cite[Chapter~6]{ManzonettoTh}, we have investigated the problem of whether the equational/order theory of a non-syntactical model 
of $\lambda$-calculus living in one of the main semantics can be r.e.
For this reason we have introduced an appropriate notion of effective model of $\lambda$-calculus, which covers in particular all the models individually 
introduced in the literature.
We have proved that the order theory of an effective model is never r.e., and hence that its equational theory cannot be $\lambda\beta$ or $\lambda\beta\eta$.
Then, we have shown that no effective model living in the stable or in the strongly stable semantics has an r.e.\ equational theory.
Concerning continuous semantics, we have investigated the class of graph models and proved that no order graph theory can be r.e., 
that many effective graph models do not have an r.e.\ equational theory and that there exists an effective graph model whose equational/order theory is the minimum one.

\bibliography{biblio}

\begin{thebibliography}{10}

\bibitem{AbramskyJM00}
S.~Abramsky, R.~Jagadeesan, and P.~Malacaria.
\newblock Full abstraction for {PCF}.
\newblock {\em Inf. Comput.}, 163(2):409--470, 2000.

\bibitem{AspertiL91}
A.~Asperti and G.~Longo.
\newblock {\em Categories, types and structures: an introduction to category
  theory for the working computer scientist}.
\newblock MIT Press, Cambridge, MA, 1991.

\bibitem{Bare}
H.P. Barendregt.
\newblock {\em The lambda calculus: its syntax and semantics}.
\newblock North-Holland, Amsterdam, 1984.

\bibitem{BastoneroTh}
O.~Bastonero.
\newblock Mod\`eles fortement stables du $\lambda$-calcul et r\'esultats
  d'incompl\'etude, 1996.
\newblock Th\`ese de {D}octorat.

\bibitem{BastoneroG99}
O.~Bastonero and X.~Gouy.
\newblock Strong stability and the incompleteness of stable models of
  $\gl$-calculus.
\newblock {\em Annals of Pure and Applied Logic}, 100:247--277, 1999.

\bibitem{BerardiB02}
S.~Berardi and C.~Berline.
\newblock $\beta\eta$-complete models for system {$F$}.
\newblock {\em Mathematical structure in computer science}, 12:823--874, 2002.

\bibitem{Berline00}
C.~Berline.
\newblock From computation to foundations via functions and application: the
  $\lambda$-calculus and its webbed models.
\newblock {\em Theoretical Computer Science}, 249:81--161, 2000.

\bibitem{Berline06}
C.~Berline.
\newblock Graph models of $\lambda$-calculus at work, and variations.
\newblock {\em Math. Struct. for Comput. Sci.}, 16:1--37, 2006.

\bibitem{BerlineMS08}
C.~Berline, G.~Manzonetto, and A.~Salibra.
\newblock Effective lambda models versus recursively enumerable lambda
  theories.
\newblock {\em Mathematical Structures in Computer Science}.
\newblock Submitted.

\bibitem{BerlineMS07}
C.~Berline, G.~Manzonetto, and A.~Salibra.
\newblock Lambda theories of effective lambda models.
\newblock In {\em Proc.\ 16th EACSL Annual Conference on Computer Science and
  Logic (CSL'07)}, volume 4646 of {\em LNCS}, pages 268--282, 2007.

\bibitem{Berry78}
G.~Berry.
\newblock Stable models of typed lambda-calculi.
\newblock In {\em Proceedings of the Fifth Colloquium on Automata, Languages
  and Programming, LNCS 62}, Berlin, 1978. Springer-Verlag.

\bibitem{BerryC85}
G.~Berry, P.-L. Curien, and J-J. L{\'e}vy.
\newblock {Full Abstraction for sequential languages: the state of the art}.
\newblock {\em Algebraic Methods in Semantics}, pages 89--132, 1985.

\bibitem{BucciarelliE91}
A.~Bucciarelli and T.~Ehrhard.
\newblock Sequentiality and strong stability.
\newblock In {\em Sixth Annual IEEE Symposium on Logic in Computer Science},
  pages 138--145. IEEE Computer Society Press, 1991.

\bibitem{BucciarelliE01}
A.~Bucciarelli and T.~Ehrhard.
\newblock On phase semantics and denotational semantics: the exponentials.
\newblock {\em Ann. Pure Appl. Logic}, 109(3):205--241, 2001.

\bibitem{BucciarelliEM07}
A.~Bucciarelli, T.~Ehrhard, and G.~Manzonetto.
\newblock Not enough points is enough.
\newblock In {\em Proc.\ 16th EACSL Annual Conference on Computer Science and
  Logic (CSL'07)}, volume 4646 of {\em LNCS}, pages 298--312, 2007.

\bibitem{BucciarelliS04}
A.~Bucciarelli and A.~Salibra.
\newblock The sensible graph theories of lambda calculus.
\newblock In {\em 19th Annual IEEE Symposium on Logic in Computer Science
  (LICS'04)}, pages 276--285. IEEE Computer Society Publications, 2004.

\bibitem{BucciarelliS0X}
A.~Bucciarelli and A.~Salibra.
\newblock Graph lambda theories.
\newblock {\em Mathematical Structures in Computer Science}, 18(5):975--1004,
  2008.

\bibitem{BurrisS81}
S.~Burris and H.P. Sankappanavar.
\newblock {\em A course in universal algebra}.
\newblock Springer-Verlag, Berlin, 1981.

\bibitem{Church32}
A.~Church.
\newblock A set of postulates for the foundation of logic.
\newblock {\em Annals of Math. (2)}, 33:346--366, 1932.

\bibitem{Comer71}
S.~Comer.
\newblock Representations by algebras of sections over boolean spaces.
\newblock {\em Pacific J. Math.}, 38:29--38, 1971.

\bibitem{DezaniLP98}
M.~Dezani-Ciancaglini, U.~de'Liguoro, and A.~Piperno.
\newblock A filter model for concurrent lambda-calculus.
\newblock {\em SIAM J. Comput.}, 27(5):1376--1419, 1998.

\bibitem{DiGianantonioFH91}
P.~{Di Gianantonio}, G.~Franco, and F.~Honsell.
\newblock Game semantics for untyped lambda calculus.
\newblock In {\em Proc. of the conference Typed Lambda Calculus and
  Applications}.

\bibitem{DiGianantonioHP95}
P.~{Di Gianantonio}, F.~Honsell, and G.~Plotkin.
\newblock Uncountable limits and the lambda calculus.
\newblock {\em Nordic Journal of Computing}, 2(2):126--145, Summer 1995.

\bibitem{Girard88}
J.-Y. Girard.
\newblock Normal functors, power series and the $\lambda$-calculus.
\newblock {\em Annals of pure and applied logic}, 37:129--177, 1988.

\bibitem{GouyTh}
X.~Gouy.
\newblock Etude des th\'eories \'equationnelles et des propri\'et\'es
  alg\'ebriques des mod\`eles stables du $\gl$-calcul, 1995.
\newblock Th\`ese, Universit\'e de Paris 7.

\bibitem{HonsellTLCA}
F.~Honsell.
\newblock {TLCA} list of open problems: Problem \# 22, 2007.
\newblock \newline\url{http://tlca.di.unito.it/opltlca/problem22.pdf}.

\bibitem{HonsellR92}
F.~Honsell and S.~Ronchi~Della Rocca.
\newblock An approximation theorem for topological lambda models and the
  topological incompleteness of lambda calculus.
\newblock {\em Journal of Computer and System Sciences}, 45:49--75, 1992.

\bibitem{Hyland75}
J.M.E. Hyland.
\newblock A syntactic characterization of the equality in some models for the
  lambda calculus.
\newblock {\em J. London Math. Soc. (2)}, 12(3):361--370, 1975/76.

\bibitem{HylandO00}
J.M.E. Hyland and C.-H.L. Ong.
\newblock On full abstraction for {PCF: I, II, and III}.
\newblock {\em Information and Computation}, 163(2):285--408, 2000.

\bibitem{Johnstone82}
P.T. Johnstone.
\newblock {\em Stone spaces}.
\newblock Cambridge University Press, 1982.

\bibitem{Kerth98b}
R.~Kerth.
\newblock Isomorphism and equational equivalence of continuous lambda models.
\newblock {\em Studia Logica}, 61:403--415, 1998.

\bibitem{Kerth01}
R.~Kerth.
\newblock On the construction of stable models of $\lambda$-calculus.
\newblock {\em Theoretical Computer Science}, 269:23--46, 2001.

\bibitem{LusinS04}
S.~Lusin and A.~Salibra.
\newblock The lattice of lambda theories.
\newblock {\em Journal of Logic and Computation}, 14:373--394, 2004.

\bibitem{MacLaneS71}
S.~{Mac Lane}.
\newblock {\em Categories for the working mathematician}.
\newblock Number~5 in Graduate Texts in Mathematics. Springer-Verlag, 1971.

\bibitem{ManzonettoTh}
G.~Manzonetto.
\newblock {\em Models and theories of lambda calculus}.
\newblock PhD thesis, Univ. Ca'Foscari (Venice) and Univ. Paris Diderot (Paris
  7), 2008.

\bibitem{ManzonettoS06}
G.~Manzonetto and A.~Salibra.
\newblock Boolean algebras for lambda calculus.
\newblock In {\em Proc.\ 21\textsuperscript{th} {IEEE} Symposium on Logic in
  Computer Science ({LICS} 2006)}, pages 139--148, 2006.

\bibitem{ManzonettoS08}
G.~Manzonetto and A.~Salibra.
\newblock Applying universal algebra to lambda calculus.
\newblock {\em Journal of Logic and Computation}, 2008.
\newblock doi: 10.1093/logcom/exn085.

\bibitem{McKenzieMT87}
R.N. McKenzie, G.F. McNulty, and W.F. Taylor.
\newblock {\em Algebras, lattices, varieties, Volume I}.
\newblock Wadsworth Brooks, Monterey, California, 1987.

\bibitem{Pierce67}
R.S. Pierce.
\newblock {\em Modules over commutative regular rings}.
\newblock Memoirs Amer. Math. Soc., 1967.

\bibitem{PigozziS98}
D.~Pigozzi and A.~Salibra.
\newblock Lambda abstraction algebras: coordinatizing models of lambda
  calculus.
\newblock {\em Fundam. Inf.}, 33(2):149--200, 1998.

\bibitem{Plotkin93}
G.D. Plotkin.
\newblock Set-theoretical and other elementary models of the lambda-calculus.
\newblock {\em Theoretical Computer Science}, 121(1\&2):351--409, 1993.

\bibitem{Salibra00}
A.~Salibra.
\newblock On the algebraic models of lambda calculus.
\newblock {\em Theoretical Computer Science}, 249:197--240, 2000.

\bibitem{Salibra01}
A.~Salibra.
\newblock A continuum of theories of lambda calculus without semantics.
\newblock In {\em 16th Annual IEEE Symposium on Logic in Computer Science},
  pages 334--343. IEEE Computer Society Press, 2001.

\bibitem{Salibra03}
A.~Salibra.
\newblock Topological incompleteness and order incompleteness of the lambda
  calculus. \uppercase{LICS}'01 \uppercase{S}pecial \uppercase{I}ssue.
\newblock Number~4, pages 379--401. ACM Transactions on Computational Logic,
  2003.

\bibitem{Scott72}
D.S. Scott.
\newblock Continuous lattices.
\newblock In {\em Toposes, algebraic geometry and logic}, Berlin, 1972.
  Springer-Verlag.

\bibitem{SelingerTh}
P.~Selinger.
\newblock Functionality, polymorphism, and concurrency: a mathematical
  investigation of programming paradigms, 1997.
\newblock PhD thesis, University of Pennsylvania.

\bibitem{Selinger03}
P.~Selinger.
\newblock Order-incompleteness and finite lambda reduction models.
\newblock {\em Theoretical Computer Science}, 309:43--63, 2003.

\bibitem{Viggo94}
V.~Stoltenberg-Hansen, I.~Lindstr\"om, and E.R. Griffor.
\newblock {\em Mathematical theory of domains}.
\newblock Cambridge University Press, New York, NY, USA, 1994.

\bibitem{Vaggione96a}
D.~Vaggione.
\newblock $\uppercase{\mathcal{v}}$ with factorable congruences and
  $\uppercase{\mathcal{v}} = \uppercase{\mathrm{i}\Gamma}^a
  (\uppercase{\mathcal{v}}_{\uppercase{di}})$ imply $\uppercase{\mathcal{v}}$
  is a discriminator variety.
\newblock {\em Acta Sci. Math.}, 62:359--368, 1996.

\bibitem{Vaggione96b}
D.~Vaggione.
\newblock Varieties in which the {Pierce} stalks are directly indecomposable.
\newblock {\em Journal of Algebra}, 184:424--434, 1996.

\bibitem{Visser80}
A.~Visser.
\newblock Numerations, $\lambda$-calculus, and arithmetic.
\newblock In Hindley and Seldin, editors, {\em Essays on Combinatory Logic,
  Lambda-Calculus, and Formalism}, pages 259--284. Academic Press, 1980.

\bibitem{Wadsworth76}
C.P. Wadsworth.
\newblock The relation between computational and denotational properties for
  {S}cott's {$D_\infty$}-models of the lambda-calculus.
\newblock {\em SIAM J. Comput.}, 5(3):488--521, 1976.

\end{thebibliography}
\bibliographystyle{plain}

\end{document}